# Revenge Porn: A Peep into its Awareness among the Youth of Tamilnadu, India

Mohammed Marzuk T.M.[1], Vijayasarathy R.[1], Madona Mathew[2]*

**ABSTRACT**
The act of posting a person's private photos or films without their consent is known as revenge porn, and it's typically done to extort money from the victim or exact retribution. According to a survey conducted by the cyber-crime in 2010, 18.3% of women didn't even know that they were victims of revenge porn. This is especially true in a country with a large population, like India, where the likelihood of this happening is even higher, but there is no specific law against revenge porn. Purposive sampling was the method utilised to choose the sample for this study. The study's sample size was 200 people, and the participants were Tamil Nadu based-unmarried women between the ages of 18 and 30. According to the results of our survey, more than 50% of people have never even heard the phrase "Revenge Porn," and only about 5% of them have ever been a victim. 40% of them believe the victim is to blame for the crime. About 43.5% of them were unsure whether or not pornographic websites should be banned. 11% of the samples indicated a likelihood that they would upload pornographic material as payback. 8.5% of respondents believe that because sex is considered taboo, society assumes that the victim is to blame for the act of retaliatory porn. Police officers need to be taught in various techniques for dealing with the victim psychologically, among other things. India, which ranks third in the world for cybercrime, may take better precautions to avoid it. People may become more aware of this crime through the implementation of particular legislation and the development of awareness campaigns.

Revenge Porn is an act of crime where sensitive content of a person is posted without their consent. We all are familiar about the term pornography yet most of the population are unaware about the happening of porn on basis of revenge, money or sometimes even for fun where illicit images or videos of women are posted in the internet or sent personally to the victim in order to blackmail. According to a survey done by the cybercrime in the year 2010. 18.3% of women didn't even know that they are victim of revenge porn. As boom in the internet world where things can reach millions in matter of seconds actions like revenge porn create huge impact on the victim as the number of viewers of the contents are not 1 or 2 its millions and its directly affect the mental health, lifestyle of the victim even though they didn't do anything wrong by law.

[1]Student, B.Sc. Forensic Science, Srinivasan College of Arts and Science, Perambalur, Tamil Nadu, India
[2]Assistant Professor - Forensic Science, Srinivasan College of Arts and Science, Perambalur, Tamil Nadu, India
*Corresponding Author

As per the survey conducted by cyber and law foundation and NGO about 27% of internet users in India aged 13 to 45 have been the victim of revenge porn. In judgmental country like India the first choice of victim is to either commit suicide or never file a complaint. Most of the times, fear of being victim shamed by others is the reason for not filing a complaint. It also starts a chain of questions and blaming the victim as why the victim shared such content with the accused in the first place. This not only breaks the victim mentally but also prevent them from taking any further steps. Further, the law does not mandate the presence female officers in such cases which creates hesitation in victims to confess to a male officer. In India, as the most populated country in the world where ratio of such crimes potentially be higher, there is no specific law against revenge porn and this serves as a main reason for people not being aware of the crime.

As example of how weak the law in India against revenge porn, there was a case in Delhi known as Air force Bal Bharati school case where a student in a class room as a revenge of his classmates calling him ugly created a site http//: www.amazing-gents.8m.net which functioned as a free space for the school students where illicit sexual descriptions of girls and teachers were posted in written form calling them with sexual names and sharing the desire to do with them. Case was filed against the student after a boy talked about the site to a girl who featured in the site. The student was sent to juvenile for a week and got released.

The punishment for spoiling the psychological health of various women is a week of jail and there is another incident in 2008 where 3 women spoke about surviving revenge porn in a blog in which only 1 out of 3 went forward to file a complaint but the officers (both male and female) mocked her by blaming her for sending such things as a result the women returned without filing a complaint. The maximum punishment for revenge porn in India is about 5 years of jail and 10 lakhs fine which is very less for a person who create a lifetime impact on the victim because things on internet can't be erased forever and that haunts the mental health of the victim throughout their life. There is a case, The state of West Bengal vs. Animesh Box, 2018 where the court treated the victim as rape survivor and gave her appropriate compensation. Things like this are required to create more awareness among people which motivates them to come out.

The main aim of the study is to analyse how many in the population are aware of the happening of revenge porn and the term revenge porn, to see the opinion of the population regarding such crime which shows us the spectrum of understanding of the population regarding the crime and to see whether the population is ready to fight against revenge porn legally. The ultimate purpose of the study is to create awareness about the crime which should nearly considered as rape but ignored by the government in most cases.

## METHODOLOGY
*Sample*
200 unmarried women of age group between 18-30 were taken as samples. The Indian state of focus for this study was Tamil Nadu. The data was collected between the period September 2022- November 2022.

# Revenge Porn: A Peep into its Awareness among the Youth of Tamilnadu, India

*Primary data*

A questionnaire form with a necessary set of questions regarding the topic was circulated among the sample and data was collected.

*The questionnaire included questions related to:*
1. Have you ever heard of the term revenge porn before?
2. Have you ever been a victim of revenge pornography threat?
3. Would you report to police if you are a victim of revenge pornography?
4. Boyfriend/Girlfriend threatens to post sexual images of you (if you are to breakup), will I you stay with them out of fear or not?
5. Would you have suicidal tendencies if you become a victim of revenge pornography?
6. Do you think victim could be somewhat responsible for triggering culprit to commit offence of revenge pornography?
7. Websites that allow people to post sexual media (For example: Myex.com, viralpop.com) are not completely banned from the internet. Do you think they should be banned?
8. Will you feel safe again with someone else after being a victim?
9. How would you react this you are one of the victims of revenge pornography?
10. Do you think looking sex as taboo makes people to blame the victim instead of the culprit?
11. If you have to decide whose responsible for revenge porn whom you think it will be?
12. Will you ever even think of harassing your partner/ex with revenge porn?
13. Being Anonymous is the Motto of using Internet, but do you think cyber-crime like revenge porn will reduce if identity of the user is revealed?
14. Do you think access to excess pornography might be the reason to influence committing revenge porn?

*Secondary data*

The secondary data was collected from digital media (Google form).

## RESULTS
### 1. Have you ever heard of the term revenge porn before?

Surprisingly, 55% of the population know about the term revenge porn and 45% haven't heard about it.

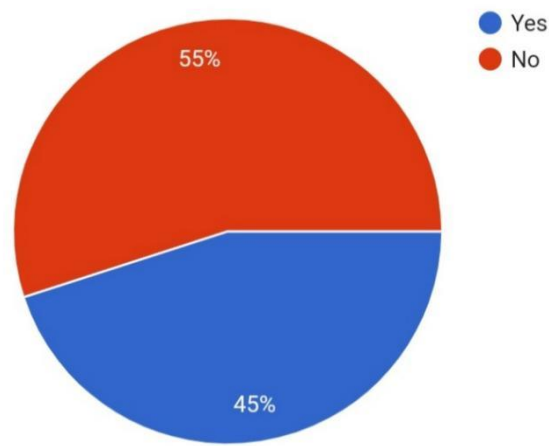

*Figure 1: Pie Chart showing the percentage of population aware of the crime*

2. **Have you ever been a victim of revenge pornography threat?**

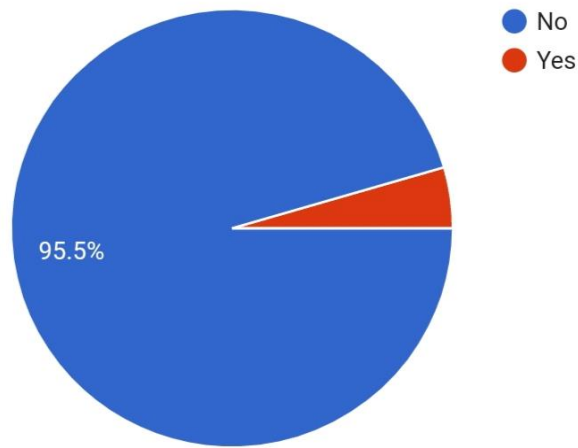

According to our survey about 4.5% of the population have been the victim of revenge porn.

*Figure 2: Pie Chart showing the percentage of population experienced the crime*

3. **Would you report to police if you are a victim of revenge pornography?**

About 81.5% of the population are more likely to file a complaint if they ever been victimized of this crime but 18.5% of the population are unlikely to file a complaint.

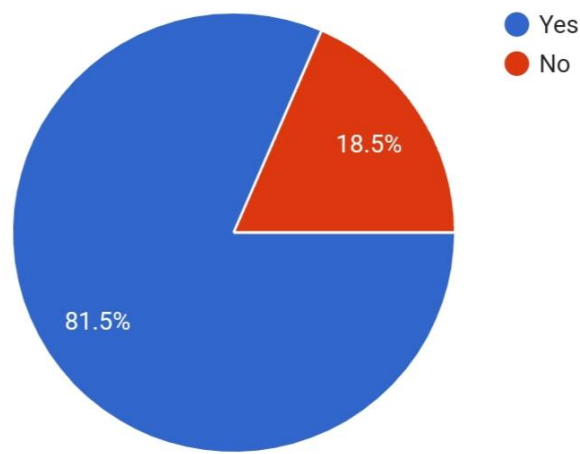

*Figure 3: Pie Chart showing the percentage of the population more likely to file a complaint if they ever be a victim*

4. **Boyfriend/Girlfriend threatens to post sexual images of you (if you are to breakup), will you stay with them out of fear or not?**

Being with the person who is blackmails with illicit images is a nightmare but surprisingly 8.5% of the population are ready to be with them out of fear.

**Revenge Porn: A Peep into its Awareness among the Youth of Tamilnadu, India**

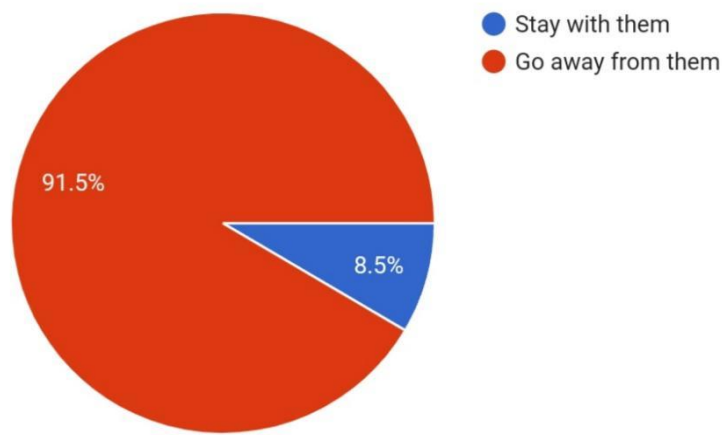

*Figure 4: Pie Chart showing the percentage of population who stay with the culprit out of fear.*

5. **Would you have suicidal tendencies if you become a victim of revenge pornography?**

Revenge porn tend to leave a huge scare in the mental health of the victims, according to our survey about 35% of the population show signs of having suicidal tendency if they ever become a victim of the crime and 3.5% of the population are not sure about how their mental state would be.

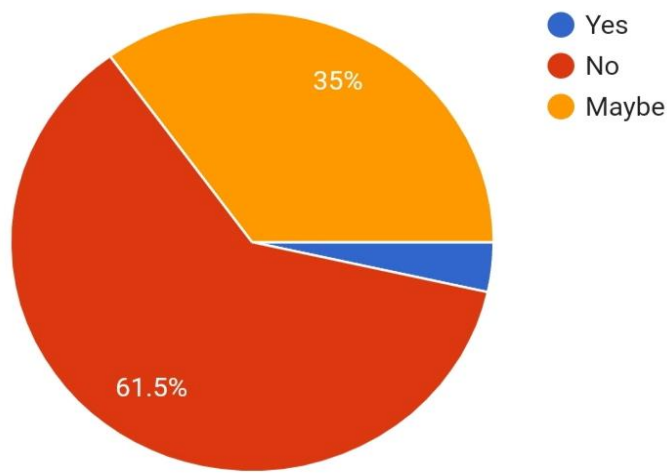

*Figure 5: Pie Chart showing the percentage of population who would have suicidal thoughts if they ever be a victim of the crime.*

6. **Do you think victim could be somewhat responsible for triggering culprit to commit offence of revenge pornography?**

Indian society is a judgmental one where blaming the victim for the crime is common and often been the reason for many to not file a complaint against the crime which definitely needs to change, our survey didn't find anything contrast to the classic Indian victim blaming, About 40% of the population consider victims might be responsible for the crime to happen, 40% person is large considering the age pool of the population is 18 to 30 which is considerably young and most of them are new gen people.

© The International Journal of Indian Psychology, ISSN 2348-5396 (e)| ISSN: 2349-3429 (p) |    212

**Revenge Porn: A Peep into its Awareness among the Youth of Tamilnadu, India**

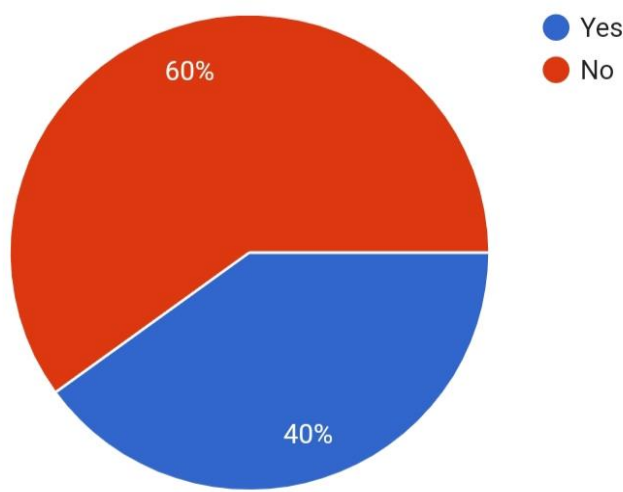

*Figure 6: Pie Chart showing the percentage of the population who consider victims might be responsible for the happening of the crime*.

7. **Websites that allow people to post sexual media (For example: Myex.com, viralpop.com) are not completely banned from the internet. Do you think they should be banned?**

There are lot of websites which provides contents exclusive based on revenge porn where people interact about these stuffs and comment on the videos posted on the site which shows how many people are out there who don't even consider the fact that the life of the women in the video have been lost due to this and ultimately lust overwhelm them.

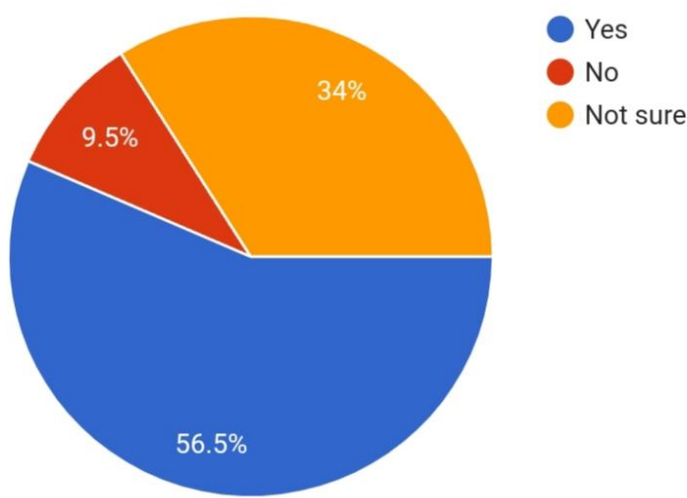

*Figure 7: Pie Chart showing the percentage of population who wants the sites posting the videos should get banned.*

As internet development had been huge recently many are aware how to open their own site and post contents they get from cheating a women. So, when the population was asked about the banning of such websites about 34% of the population are not sure about banning of those websites.

### 8. Will you feel safe again with someone else after being a victim?

Having your sexual image shared in internet among of millions of internet users would create a huge impact on the psychological health of a victim. Many tend to disconnect from others mostly because a image is formed inside them convincing everyone out there tend to do the same thing if they ever have to be with someone after the occurrence of the crime. Psychological support need to be given to the victims making them believe that this is not end of the world. So, unsurprisingly about 81.5% of the population are sure that they won't feel safe with someone after being the victim of the crime.

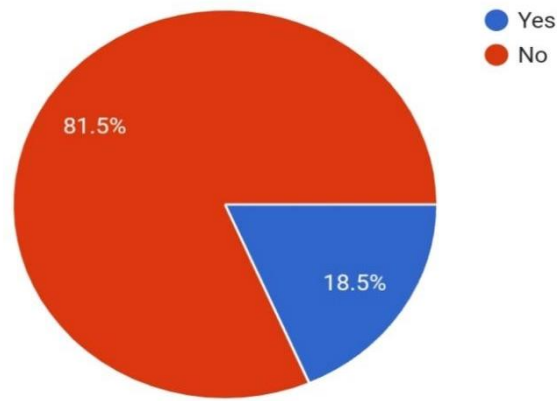

*Figure 8: Pie Chart showing the percentage of population who wont be safe with someone else after being the victim of the crime.*

### 9. How would you react this you are one of the victim of revenge pornography?

Awareness of the crime among the people in India is very less and even in the case them being aware about this, moving things legally is a big question mark as people in the society tend to see things like this as taboo and blame the victim for the happening of the crime. According to a survey conducted by cyber and law foundation and NGO about 35% of the women whose illicit stuffs have been released are tend not to file a complaint in India. So, when our population had the question of how they will react if they ever become a victim put forward, 81.5% of them consider filing a complaint legally which is a great sign and 16% of the population are not sure how they might react to the situation and 2.5% of the population are sure about them not filing any complaint.

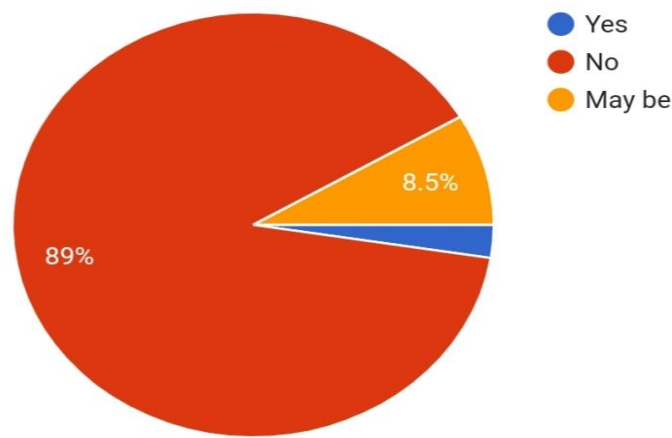

*Figure 9: Pie Chart showing the percentage of population on how they will react if they ever become a victim of the crime.*

**10. Do you think looking sex as taboo makes people to blame the victim instead of the culprit?**

Sex education is considered taboo even in 2023 in India, there are schools even now where the lesson reproduction is either skipped or not taught completely. As we grew up not once listening anything about sex in our life from our parents or teachers there is a stigma in our mind that even if I ever come out bravely as victim of the crime people will see me as taboo. So, about 68.5% of the population consider seeing sex as taboo is a reason why society tend to blame the victim often instead of the culprit.

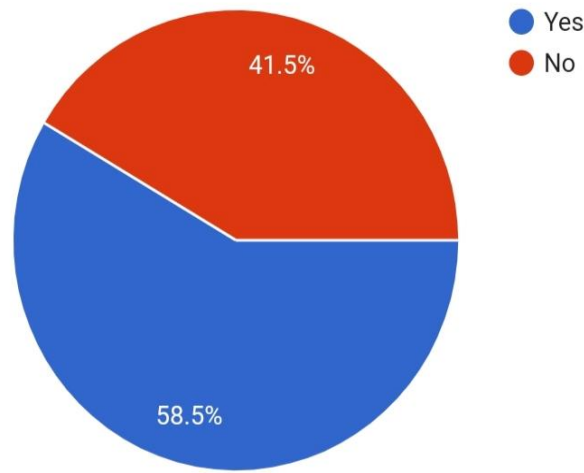

*Figure 10: Pie Chart showing the percentage of population who consider seeing sex as taboo is the reason to blame victim.*

**11. If you have to decide whose responsible for revenge porn whom you think it will be?**

Whatever may be the situation haunting a person by posting their nudes in online is not an acceptable option. Yes, the power of committing the crime is given by the victim but anyone in right mind wont misuse them and catfishing just to get the illicit photos or videos are nothing less than planned rape. So, 45.5% of the population consider both the victim and the culprit are responsible for the occurrence of the crime, 46% of the population consider doer is responsible for the crime and 8.5% of the population consider victim is responsible for the crime which shows there are still people in this young generation who blame the victim.

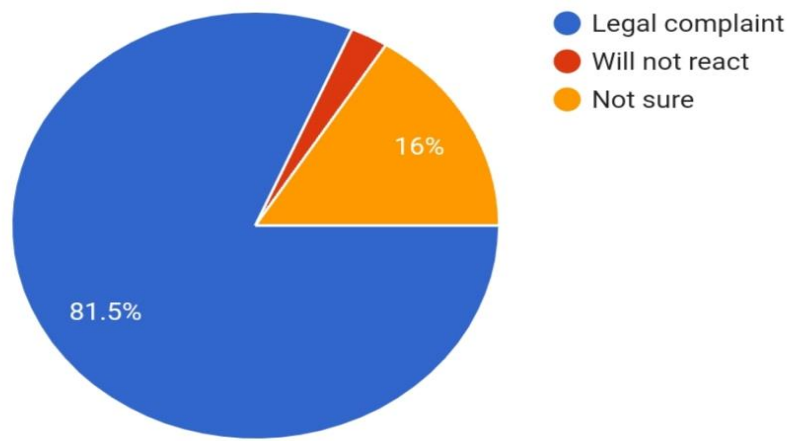

*Figure 11: Pie Chart showing the opinion of the population on who is responsible for the crime.*

**12. Will you ever even think of harassing your partner/ex with revenge porn?**

Revenge porn is often ultimate result of toxic relationships, on the study done by Mount Sinai, about 1 in 3 teens have been either verbally or physically abused by their partners. Not ending things in the start of such relationships might ruin someone's whole life. So, when we tested the potential toxic thoughts of the population about 8.5% of the population are sure that they thought about taking revenge on their partner or ex with revenge porn.

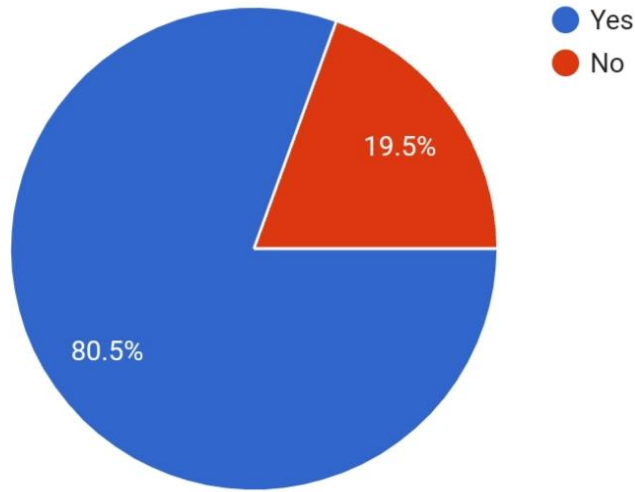

*Figure 12: Pie Chart showing the percentage of population who consider taking revenge with revenge porn.*

**13. Being Anonymous is the Moto of using Internet, but do you think cybercrime like revenge porn will reduce if identity of the user is revealed?**

Ultimate aim of the internet is to provide anonymity and many tend to misuse the concept of it. You can see gore contents in dark web which will be out of your expectation yet government cant take actions on those things because simply they don't know who is responsible for it. So, people hide behind this big anonymity concept which brings the space transition theory in play according to which people complete the desire in internet which they cant do in real life and its possible because of the anonymity provided by the internet. 80.5% of the population consider that the rate crime might reduce if the identity of the user is revealed in internet.

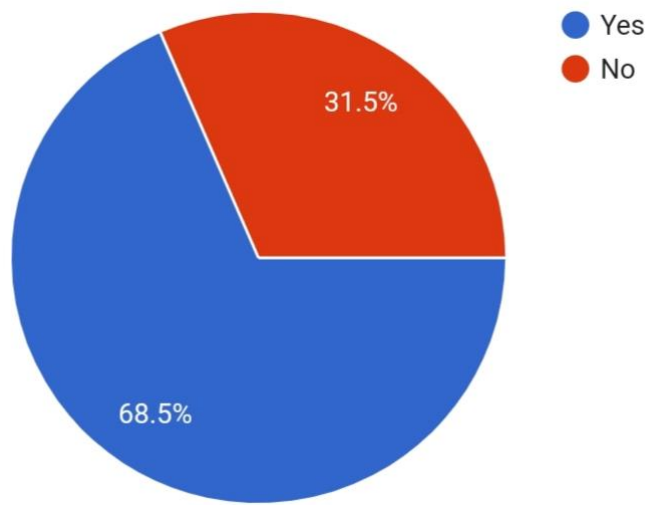

*Figure 13: Pie chart showing the percentage of population who consider revealing the identity of the user might reduce rate of the crime.*

**14. Do you think access to excess pornography might be the reason to influence committing Revenge porn?**

Access to pornography is no big thing at a moment. Actually, the access become so much easier nowadays as the transition from google to telegram or twitter happened.

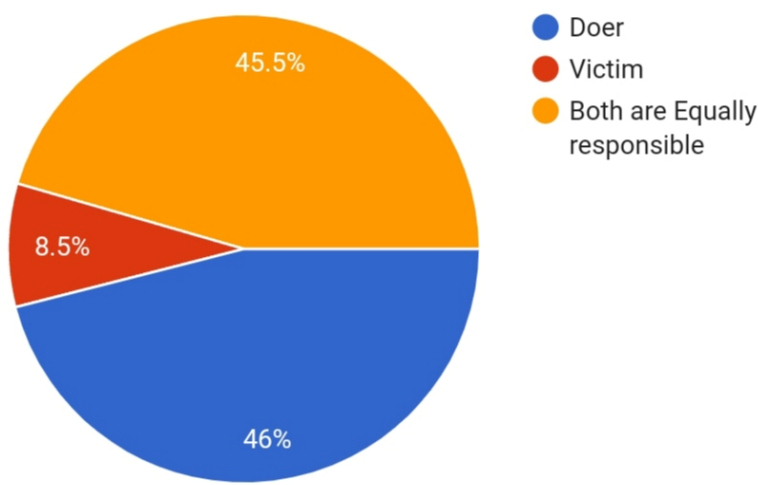

*Figure 14: Pie Chart showing the percentage of population who consider excess access to pornography as influence to commit such crimes.*

There are many dedicated pages in telegram and twitter which post porn content for free. We become what we consume, about 58.5% of the population consider excess access to pornography might be an influence in committing such crimes.

**GOVERNMENT INITIATIVES**

Revenge Pornography still hasn't recognised by everyone in India considering there isn't much awareness about it among the people so there is not any specific laws for revenge Pornography in India. Cases of revenge Pornography is being mostly filed under sexual harassment and breach of privacy.

Some of the Acts present in India for sharing non-consensual illicit contents comes under:

*Information technology Act 2000*
IT Act includes punishment of upto ten years or fine or with both,
- Section 66E – violation of privacy of victims by capturing images of private sexual body parts without their permission.
- Section 67 and 67A – creating and sharing of illicit sexual content through electronic medium.
- Section 67B – This section includes any transmission of content of victims who are under the age of 18.
- Section 72 – Breach of Confidentiality and privacy

*Indian Penal Code*
- Section 292 – selling or publication of obscene materials
- Section 354 and 354A – Criminal force or assault on women for outraging modesty and man who advance on women for physical contact without context or forces her are punished with 3 years and liable to fine.

- Section 354C – Man who watches or captures a women engaging in a private act without her permission is committing voyeurism.
- Section 509 – Man who insults a women by words, sounds or gestures is punishable under 1 year or fine or both.

*The Indecent Representation of Women (Prohibition) Act*
- Section 4 – Prohibits sending and sharing of paper, slide, painting or photography which contains indecent information of women.
- Section 6 – Provides vigorous punishment for violating section 4

*Cyber cells*
Cybercells are present in every state and union territories. These Cybercell websites can be accessed by web browser and can be used to file any type of Cyber Crime such as Revenge Porn.

# CONCLUSION
As per our findings, we concluded that Revenge pornography has been growing in India even though half of the population doesn't even know about the term. A survey done by UK Charity Refugee in 2020 revealed 27% of men and 43% of women share intimate images when being in a relationship. Sharing of sexual content in a relationship ends up being a risk cause they can be distributed in motives of revenge. Another survey by UK Charity Regugee found that that in fourteen adults received threats on having their sexual content published. UK has updated their online safety bill on 2022 to prevent websites and companies from sharing illicit content and if they are to violate, they need to pay 10% of their annual income as fine. Indian laws are very similar to UK's law so it would be better if such laws like online the safety bill and specific laws for revenge pornography are introduced. Even when search Engines like GOOGLE are trying to remove these kinds of content from the web, they can't be removed from the digital media permanently. In states like Tamilnadu, awareness of these crimes are less and they tend to trust information technology. The government should take necessary actions to provide people with necessary information on cybercrimes and its laws.

*Acknowledgement*
The author(s) appreciates all those who participated in the study and helped to facilitate the research process.

*Conflict of Interest*
The author(s) declared no conflict of interest.

*How to cite this article:* Mohammed, M.T.M., Vijayasarathy, R., & Mathew, M. (2023). Revenge Porn: A Peep into its Awareness among the Youth of Tamilnadu, India. *International Journal of Indian Psychology*, *11(3),* 208-219. DIP:18.01.018.20231103, DOI:10.25215/1103.018